\documentclass[prl,floatfix,showpacs,showkeys,twocolumn]{revtex4}
\usepackage{graphicx}
\begin{document}

\title{
   Residual Interactions and Correlations 
   Among Laughlin Quasiparticles:\\
   Novel Hierarchy States}

\author{
   John J. Quinn$^{a,*}$, 
   Arkadiusz Wojs$^{a,b}$, 
   and Kyung-Soo Yi$^{a,c}$}

\affiliation{
   $^a$University of Tennessee, Knoxville, Tennessee 37996, USA\\
   $^b$Wroclaw University of Technology, 50-370 Wroclaw, Poland\\
   $^c$Pusan National University, Busan 609-735, South Korea}

\begin{abstract}
The residual interactions between Laughlin quasiparticles can be obtained 
from exact numerical diagonalization studies of small systems.
The pseudopotentials $V_{\rm QP}(\mathcal{R})$ describing the energy of
interaction of QE's (or QH's) as a function of their ``relative angular
momentum'' $\mathcal{R}$ cannot support Laughlin correlations at certain 
QP filling factors (e.g., $\nu_{\rm QE}=1/3$ and $\nu_{\rm QH}=1/5$).
Because of this  the novel condensed quantum fluid states observed at 
$\nu =4/11$, 4/13 and other filling fractions cannot possibly be spin 
polarized Laughlin correlated QP states of the composite Fermion hierarchy.
Pairing of the QP's clearly must occur, but the exact nature of the
incompressible ground states is not completely clear.
\end{abstract}



\maketitle

Fractional quantum Hall states have been observed recently at unexpected 
values of the electron filling factor $\nu$ \cite{pan03}.
Some of these states have been attributed to composite Fermions (CF's) 
of different ``flavor'' with the notation $^2$CF, $^4$CF, \dots used  
for CF's with different numbers of attached Chern--Simons (CS) flux quanta
\cite{pan03,smet03}.
This idea is not new.
It is equivalent to a CF hierarchy scheme \cite{sitko97}, which involved the
reapplication of the CS transformation to quasiparticles (QP's) in partially
filled CF angular momentum shells (or Landau levels) proposed to describe odd
denominator fractions that did not belong to the Jain sequence \cite{jain89} 
of filling factors.
Furthermore, it is known from exact numerical diagonalization studies of 
small system that certain fractional filling (e.g. $\nu=4/11$ corresponding 
to quasielectron (QE) filling fraction $\nu_{\rm QE}=1/3$ and $\nu=4/13$ 
corresponding to quasihole (QH) filling $\nu_{\rm QH}=1/5$) do not possess 
Laughlin-type incompressible liquid ground states \cite{sitko97,quinn00-pe}.
The reason for this is that the CS transformation applied to QP's in the CF
hierarchy picture is applicable only to
interacting systems which support Laughlin correlations \cite{quinn00-pe}.
By Laughlin correlations we mean the maximum avoidance of pair states with 
the largest pair angular momentum $L'$ (or smallest value of the `relative 
angular momentum', $\mathcal{R}=2l - L'$, where $l$ is the angular momentum 
of the individual particles).
In order to support Laughlin correlations 
\cite{quinn00-pe,quinn00-jpc,wojs99-ssc}, the pseudopotential $V(L')$ 
describing the interaction energy of a pair of particles as a function
of the pair angular momentum $L'$, must increase, approaching the avoided 
value of $L'$, more quickly than $L'(L'+1)$.
We refer to such a potential as ``superharmonic'' since it increases more
quickly than any $V_{\rm H}(L')=A+B\hat{L'}^2$
(where $A$ and $B$ are constants), defined as a
harmonic pseudopotential \cite{quinn00-pe,quinn00-jpc,wojs99-ssc}.
For electrons in the lowest Landau level ($n=0$), $V_0(L')$ is 
``superharmonic'' at every value of $L'$.
For excited Landau levels \cite{wojs01-prb} ($n \ge 1$) $V_n(L')$ is not
superharmonic at all the allowed values of $L'$.
Neither is the pseudopotential $V_{\rm QP}(L')$
\cite{quinn00-pe,quinn00-jpc,wojs99-ssc,wojs01-prb,sitko96,lee01}, 
describing the interaction of Laughlin quasiparticles (QP's), 
superharmonic at all allowed values of $L'$.
In these situations the interacting particles tend to form pairs in 
order to lower the total energy \cite{wojs01-prb}.
These pairing correlations can also lead to a nondegenerate incompressible
ground state.
Moore and Read \cite{moore91} proposed such an incompressible ground 
state of pairs to explain the observation of the fractional quantum Hall 
effect at $\nu=5/2$. 
For Laughlin QP's of the $\nu =1/3$ state, it has been shown that 
$V_{\rm QE}(L')$ is not superharmonic at $\mathcal{R}=1$, and 
$V_{\rm QH}(L')$ is not at $\mathcal{R}=3$.
Therefore, suggestions \cite{pan03,smet03} that $\nu=4/11$ and $\nu=4/13$ 
are daughter states in a spin polarized system that arise from Laughlin 
condensation of QP's at $\nu_{\rm QE} =1/3$ and $\nu_{\rm QH}=1/5$ cannot 
possibly be correct.

The object of the present paper is to demonstrate by both analytical and
numerical techniques that Laughlin correlations will not occur for the 
lowest energy states in the spectrum if the pseudopotential is subharmonic.
By using the quasiparticle pseudopotentials $V_{\rm QP}(L')$ obtained by
numerical diagonalization of small systems of electrons, we have obtained 
the energy spectra of systems containing a small number of QP's 
(with $4 \le N_{\rm QP} \le 18$ at QP filling factors in the range 
$1/3 \le \nu_{\rm QP} \le 2/3$).
These results are {\em thought of} as ``numerical experiments'' with which 
intuitive physical models are to be compared.
The simple models that we have considered are based on the idea that only 
two coefficients $V_{\rm QP}(\mathcal{R})$ of the QP pseudopotential play 
an important role in determining the nature of the correlations 
($\mathcal{R}=1$ and $\mathcal{R}=3$ for QE's; $\mathcal{R}=3$ and 
$\mathcal{R}=5$ for QH's, with $V_{\rm QH}(\mathcal{R}=1) \gg V_{\rm QH}$ 
at $\mathcal{R}=3$ and $\mathcal{R}=5$).
Though no simple model exactly fits the numerical experiments, it seems 
clear from the numerical experiments alone that the correlations among
the QP's which give rise to the novel fractional quantum Hall (FQH) states 
are of a new type that involves formation of pairs.
These correlations are very different from the Laughlin correlations, 
which give rise to the standard CF hierarchy of spin polarized FQH states.

To eliminate boundary conditions but preserve translational symmetry 
in a two dimensional (2D) electron gas of finite size, it has become 
customary to confine the electrons to a spherical surface of radius $R$.
A magnetic monopole of strength $2Q\phi_0$ (where $\phi_0=hc/e$ is the 
flux quantum and $2Q$ is an integer) at the center produces a radial 
magnetic field of magnitude $B=2Q\phi_0/4\pi R^2$.
The single particle eigenfunctions in this Haldane geometry \cite{haldane88},
are called monopole harmonics and denoted by $\left|Q,l,m\right>$, where 
$Q$ is half the monopole strength, $l$ the angular momentum, and $m$ its 
$z$-component.
The single particle eigenvalues are given by  
$\epsilon_l = (\hbar \omega_c/2Q)\left[l(l+1) -Q^2 \right]$,
where $\omega_c$ is the cyclotron frequency.
Because $\epsilon_l$ must be positive, the minimum value of $l$ is $Q$, 
and we can label the angular momentum shells by $l_n = Q+n$,
where $n$ is a non-negative integer.
For convenience of notation we will write the monopole harmonics as 
$\left|l,m\right>$ with $Q$ being understood.

For a system of $N$ electrons confined to a shell of angular momentum $l$, 
we can form $N$ electron eigenfunctions with a given value of $L$, the total
angular momentum, and $M$, its $z$-component.
They can be written $\left|L,M,\alpha\right>$ with the label $\alpha$ 
distinguishing distinct multiplets with the same values of $L$.
The Wigner--Eckart theorem states for a scalar interaction $H'$ that
$
\left<L',M',\alpha'\right| H'\left| L,M,\alpha\right> = \delta_{LL'}
\delta_{MM'}\left<L\alpha'\right| H'\left| L\alpha\right>$
and that the reduced matrix element on the right hand side is independent 
of $M$.
The eigenfunction for the $\alpha^{th}$ multiplet of total angular 
momentum $L$ formed by adding the angular momenta $l_i=l$ of $N$ 
identical Fermions can be written
\begin{eqnarray}
&&\left| l^N;L\alpha\right> =  \nonumber \\
&&
\sum_{L_{12}L''\alpha''}G_{L\alpha,L''\alpha''}(L_{12})
\left| l^2,L_{12};l^{N-2},L''\alpha'';L\right>.
\label{N fermion state}
\end{eqnarray}
Here the $G_{L\alpha,L''\alpha''}(L_{12})$ are `coefficients of fractional 
grandparentage' \cite{cowan}.
The wavefunctions on the right hand side of Eq.(\ref{N fermion state}) are
obtained by adding the angular momentum $L_{12}$ of the pair $<1,2>$ to the
angular momentum $L''$ of the $\alpha''$ multiplet of the $j=3$, 4, \dots, 
$N$ remaining Fermions to obtain the total angular momentum $L$.
Although $\left| l^2,L_{12};l^{N-2},L''\alpha'';L\right>$ is not
antisymmetric under the interchange of $i=1$ or 2 with $j=3$, 4, \dots, 
$N$, the eigenfunctions $\left|l_N;L\alpha\right>$ are totally antisymmetric.
We define the `pair amplitude' $\mathcal{G}_{L\alpha}(L')$ by
\cite{halperin83} $\mathcal{G}_{L\alpha}(L')=\sum_{L''\alpha''}
|G_{L\alpha,L''\alpha''}(L')|^2$.
Orthonormality of the eigenfunctions $\left|l^N;L\alpha\right>$ gives 
the sum rule
\begin{equation}
\sum_{L'} \mathcal{G}_{L\alpha}(L') =1.
\label{sum rule-1}
\end{equation}
A second useful sum rule 
\begin{eqnarray}
\frac{1}{2}N(N-1)\sum_{L'} L'(L'+1)\mathcal{G}_{L\alpha}(L')=&&
\nonumber\\
L(L+1)+N(N-2)l(l+1)&&
\label{sum rule-2}
\end{eqnarray}
can be obtained by using Eq. (\ref{N fermion state}) together with the
simple theorem on pair angular momenta 
$\hat{L}^2 + N(N-2)\hat{l}^2 = \sum_{\left<i,j\right>}
(\hat{l_i}+\hat{l_j})^2$ \cite{wojs99-ssc}.
In this equation $\hat{l_i}+\hat{l_j}$ is the angular momentum operator 
of the pair $\left<i,j\right>$, and the sum is over all pairs.
The energy of the multiplet $\left| L\alpha\right>$ is given by 
\begin{equation}
E_{\alpha}(L) = \frac{1}{2}N(N-1)\sum_{L'}\mathcal{G}_{L\alpha}(L')V(L'),
\label{energy L}
\end{equation}
where $V(L')$ is the pseudopotential.
It is clear from Eq. (\ref{energy L}) and the sum rules [Eqs. 
(\ref{sum rule-1}) and (\ref{sum rule-2})] that, for a ``harmonic 
potential'' $V_{\rm H}(L')$, the energy is given by $E_{\alpha}(L) = 
c_1 + c_2 L(L+1)$ where $c_1$ and $c_2$ are independent of $\alpha$.
Because the right hand side of this equation is independent of $\alpha$,
every multiplet with the same value of $L$ is degenerate,
and the harmonic pseudopotential introduces no
correlations \cite{quinn00-jpc,wojs99-ssc}.
Any linear combination of eigenfunctions with the same value of $L$ 
(i.e., $\sum_\alpha c_\alpha \left| L\alpha\right>$) has the same energy.

Since $\mathcal{R}=2l-L'$, we can think of the pseudopotential as 
a function of $\mathcal{R}$, and write $V(\mathcal{R}) = V_{\rm H}
(\mathcal{R})+\Delta V (\mathcal{R})$.
Correlations are completely determined by the anharmonic part 
$\Delta V(\mathcal{R})$. 
For a simple model in which $\Delta V=\triangle_1\delta_{\mathcal{R},1}$,
with the constant $\triangle_1 > 0$, the lowest energy state for each 
value of $L$ is the one with the smallest value of
$\mathcal{G}_{L\alpha}(\mathcal{R}=1)$, which we will call
$\mathcal{G}_{L0}(\mathcal{R}=1)$.
This is exactly what we mean by Laughlin correlations.
In fact, if $\triangle_1$ is infinite, the only states with finite 
energy are those for which $\mathcal{G}_{L0}(\mathcal{R}=1)$ vanishes.
The complete avoidance of the pair states with $\mathcal{R}=1$ corresponds
exactly to the Laughlin--Jastrow factor $\prod_{\left<i,j\right>}
(z_i-z_j)^2$ in the Laughlin wavefunction for the $\nu=1/3$ state 
\cite{laughlin83}.

Now let's consider a model pseudopotential which can be superharmonic or
subharmonic at $\mathcal{R}=1$, viz., one in which 
$\Delta V(\mathcal{R}) = \triangle_1 \delta_{\mathcal{R},1} 
+ \triangle_3 \delta_{\mathcal{R},3}$.
We assert that if $\triangle_3$ is sufficiently large, Laughlin correlations 
will not produce the lowest energy state. 
We demonstrate this as follows:

i) the Laughlin correlated $L=0$ ground state which occurs at $2Q=3(N-1)$ 
when $\triangle_3=0$ must have the minimum possible value of
$\mathcal{G}_0(\mathcal{R}=1)$.

ii) in the presence of $\triangle_3 >0$, decrease $\mathcal{G}_0
(\mathcal{R}=3)$ by an amount $\Delta \mathcal{G}$.

iii) in order to satisfy the first sum rule, Eq.(\ref{sum rule-1}), 
other pair amplitudes will have to increase.

\noindent
For simplicity, let's assume that only $\mathcal{G}(\mathcal{R}=1)$ and
$\mathcal{G}(\mathcal{R}=j)$, with $j$ an odd integer between $2l$ and 5,
increase.
By taking $\Delta\mathcal{G}(\mathcal{R}=1)=x_j \Delta\mathcal{G}$ and 
$\Delta\mathcal{G}(\mathcal{R}=j)=(1-x_j) \Delta\mathcal{G}$
along with $\Delta\mathcal{G}(\mathcal{R}=3)=- \Delta\mathcal{G}$,
the first sum rule is automatically satisfied.
The second sum rule, Eq.(\ref{sum rule-2}), determines $x_j$, giving 
$x_j = 1-2(4l-3)(4l-j)^{-1}(j-1)^{-1}$.
The change in energy of the $L=0$ ground state in the presence of 
$\triangle_1$ and $\triangle_3$ is given by 
\begin{equation}
\Delta E_0 = \Delta\mathcal{G}(x_j\triangle_1-\triangle_3).
\label{delta E}
\end{equation}
This becomes negative when $\triangle_3 > x_j \triangle_1$.
For example, if we take $j=5$, $x_5 = (4l-7)(8l-10)^{-1}$.
The value of $\triangle_3 = x_5\triangle_1$ is exactly the same value 
that causes $\Delta V(\mathcal{R})$ to behave harmonically between 
$\mathcal{R}=1$ and $\mathcal{R}=5$. 
It always gives a superharmonic pseudopotential $V(\mathcal{R})$ at
$\mathcal{R}=3$, but at $\mathcal{R}=1$, it is superharmonic only if 
$\triangle_3 < x_1 \triangle_1$.
It is not difficult to see that transfer of $\Delta\mathcal{G}$ to
$\mathcal{R}=1$ and $\mathcal{R}=5$ results in the minimum value of $x_j$.
The transfer of pair amplitude to pair states with $\mathcal{R}=1$ together
with the decrease in pair amplitude at $\mathcal{R}=3$ is a clear indication 
of the formation of Fermion pairs with $\mathcal{R}=1$ and the avoidance 
of pair states with $\mathcal{R}=3$ (and the maximum repulsive interaction).
Numerical studies \cite{wojs01-prb,quinn03-pla,wojs03-appa} of small systems
clearly support this picture when the pseudopotential is not superharmonic.

The number of electrons required in order to have a system of QP pairs 
of reasonable size is, in general, too large for exact diagonalization 
in terms of electron states and the Coulomb pseudopotential \cite{mandal}.
However, by restricting our consideration to the QP's in the partially 
filled CF shell, and by using the QP pseudopotential obtained from 
numerical studies \cite{quinn00-pe,quinn00-jpc,wojs99-ssc,wojs01-prb,%
sitko96,lee01} of small systems of electrons, we can reduce the numerical 
diagonalization to manageable size \cite{lee02}.
The QP pseudopotentials determined in this way \cite{quinn03-pla} are quite
accurate up to an overall constant which has no effect on the correlations.
Furthermore, because the correlations are primarily determined by the short
range part of the pseudopotential, the numerical results for small systems
should describe the essential correlations quite well for systems of any size.
\begin{figure}
\resizebox{3.40in}{4.60in}{\includegraphics{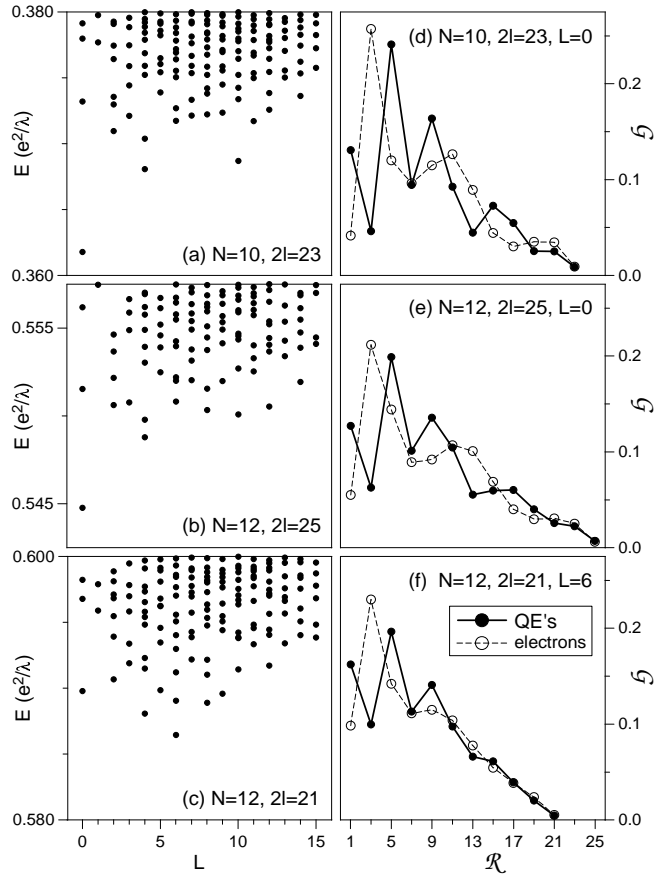}}
\caption{
Low energy spectra and pair amplitude functions:
Frames (a), (b), and (c) show the energy spectra for $N=10$ QE's at $2l=23$, 
for $N=12$ QE's at $2l=25$, and for $N=12$ QE's at $2l=21$ as a function of
total angular momentum $L$.
Frames (d), (e), and (f) display pair amplitude functions
$\mathcal{G}(\mathcal{R})$ for the ground states of the case presented 
in (a), (b), and (c), as a function of relative pair angular momentum 
$\mathcal{R}$.
The solid circles are the ground state values of $\mathcal{G}(\mathcal{R})$ 
for the QE pseudopotentials.
The open circles are the values for the superharmonic electron 
pseudopotential.
All spectra were obtained using $V_{\rm QE}(\mathcal{R})$ given 
in Ref.~\cite{lee01}.}
\label{fig1} 
\end{figure}
In Fig. 1, we present low energy spectra for three different cases: 
(a) is for $N=10$ QE's at $2l=23$, and corresponds to $\nu_{\rm QE}=1/3$ 
and $\nu=4/11$;
(b) is for $N=12$ QE's at $2l=25$, and and corresponds to $\nu_{\rm QE}
=1/2$ and $\nu=3/8$; (c) is for $N=12$ QE's at $2l=21$, and it should 
also correspond to $\nu_{\rm QE} =1/2$ and $\nu=3/8$.
The pseudopotentials given by Lee {\sl et al.} \cite{lee01} were used
in obtaining these results.
For small values of $\mathcal{R}$, their $V_{\rm QE}(\mathcal{R})$
agrees reasonably well with our earlier results \cite{quinn00-pe,%
quinn00-jpc,wojs99-ssc,wojs01-prb,sitko96}, and the spectra and
pair amplitudes are not very sensitive to which of these different
$V_{\rm QE}(\mathcal{R})$ is used.
The $\nu_{\rm QE}=1/3$ state is one of a sequence of states occurring 
at $2l=3N-7$ whose spectra we have evaluated numerically for $4 \leq N 
\leq 12$.
The other two states belong to the sequence $2l=2N+1$, which together 
with their conjugate states at $2l=2N-3$ (obtained by replacing $N$ by 
$2l+1-N$, the number of QH's) correspond to $\nu_{\rm QE} = 1/2$ and 
$\nu=3/8$.
Frames (a) and (b) show $L=0$ ground states separated by a substantial 
gap from excited states.
Frame (c) does not have an $L=0$ ground state, though a simple pairing 
model \cite{quinn03-pla,wojs03-appa} would predict one for this case.
In frames (d), (e), and (f) the values of the pair amplitude functions
$\mathcal{G}(\mathcal{R})$ as a function of $\mathcal{R}$ for the ground 
states of (a), (b), and (c) are shown as solid dots.
For the sake of contrast, $\mathcal{G}(\mathcal{R})$ for a superharmonic
electron pseudopotential are shown as open circles.
The pairing at $\mathcal{R}=1$ and avoidance of $\mathcal{R}=3$ QP states
are quite clear.

A very simple pairing model was presented \cite{quinn03-pla, wojs03-appa}
earlier which assumed that all the QE's formed $\mathcal{R}=1$ pairs.
The pairs can be treated as Bosons \cite{wojs03-appa} or as Fermions
\cite{quinn03-pla}, and if Laughlin correlations between the pairs are
assumed, incompressible ground states are formed at $\nu_{\rm QE} = 1/3$, 
1/2, and 2/3 and $\nu_{\rm QH} = 1/5$, 1/4, and 2/7 giving novel condensed 
states at the values $\nu=5/13$, 3/8, 4/11, and $\nu=5/17$, 3/10, 4/13 
observed experimentally \cite{pan03}.
However, the simple ``complete pairing'' model is probably too simple.
Two major difficulties are not yet understood.
First, the states obtained in our numerical calculations occur at 
$2l=3N-7$ (for $\nu_{\rm QE}=1/3$) for $N=8$, 9, 10, 11, and 12, and at
$2l=\frac{3}{2}N+2$ (for $\nu_{\rm QE}=2/3$) for $N=10$, 12, 14, 16, and 18.
Complete pairing can only occur for $N$ even, and the sequence at
$2l=3N-7$ occurs for both odd and even values of $N$.
In addition, the simple ``complete pairing'' model would predict the 
$\nu_{\rm QE}=1/3$ state at $2l=3N-5$ and the $\nu_{\rm QE} = 2/3$ 
state at $2l=\frac{3}{2}N+1$, instead of at the values of $2l$ observed
in the numerical study.
Although this discrepancy is a finite size effect which becomes negligible 
for large $N$, we consider it important and are trying to understand its 
cause.
It is worth noting that the formation of Fermion triplets (i.e., three 
QE's forming a compact droplet with angular momentum $3l-3$) would lead 
to the relation $2l=3N-7$, as would partial pairing with $N_1 = 
\frac{1}{3}N$ unpaired and $2N_2 = \frac{2}{3}N$ paired QE's.
However, both cases require $N$ to be divisible by three.
We are currently exploring these and other extensions of the simple model 
of complete pairing, but have no clear answer at present.
The second problem is that the $\nu_{\rm QE}=1/2$ states, which occur at
$2l=2N-3$ and $2l=2N+1$ values predicted by the simple model, are found 
in our numerical calculations as conjugate pairs at $2l=9$ and $N=4$ or 6, 
at $2l=17$ and $N=8$ or 10, and at $2l=25$ and $N=12$ or 14.
However, incompressible states are found numerically neither at $2l=13$ 
and $N=6$ or 8, at $2l=21$ and $N=10$ or 12, nor at $2l=29$ and $N=14$ 
or 16, where the simple model suggests they should occur.
\begin{figure}
\resizebox{3.40in}{3.33in}{\includegraphics{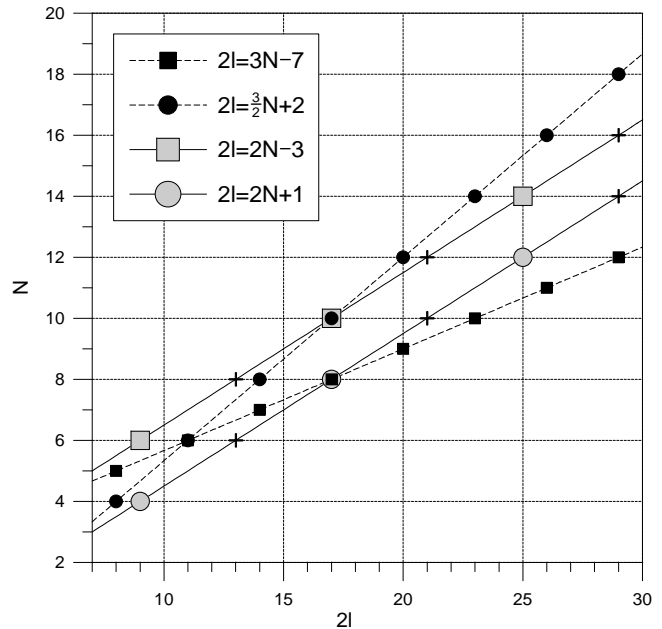}}
\caption{
The sequences $2l=3N-7$ ($\nu_{\rm QE}=1/3$) and $2l=\frac{3}{2}N+2$ 
($\nu_{\rm QE}=2/3$) and the conjugate pairs at $\nu_{\rm QE}=1/2$ 
($2l=2N-3$ and $2l=2N+1$) are shown as straight lines.
The values of $N$ and $2l$ at which $L=0$ ground states separated from 
excited states by a substantial gap are shown as solid dots and solid 
squares (for $\nu_{\rm QE}=1/3$ and 2/3, respectively) and by open 
circles and open squares (for $\nu_{\rm QE}=1/2$).
The locations where $L=0$ ground states of $N$ QP's each with angular 
momentum $l$ would be expected in the simple pairing model but are not 
found numerically are indicated by the symbol ``$+$''.}
\label{fig2} 
\end{figure}
These results are summarized in Fig. 2, a plot of $N$ versus $2l$ which 
contains four straight lines $2l=3N-7$, $2l=\frac{3}{2}N+2$, $2l=2N-3$, 
and $2l=2N+1$.
The last two are conjugate pair states for $\nu_{\rm QE}=1/2$.
The value at which $\nu_{\rm QE}=1/3$ and $\nu_{\rm QE}=2/3$ states found 
in our `numerical experiments' are shown as solid squares and solid dots, 
respectively.
The values at which we find $\nu_{\rm QE}=1/2$ states are shown as open
circles and squares (the circles and squares surround the solid dots and 
solid squares at $2l=17$, where $\nu_{\rm QE}=1/2$ and $\nu_{\rm QE}=1/3$ 
or $\nu_{\rm QE}=2/3$ fit the observed states).
The expected but unobserved states at $2l=13$ (for $N=6$ and 8), $2l=21$ 
(for $N=10$ and 12), and $2l=29$ (for $N=14$ and 16) are indicated by 
the symbol ``$+$''.
It would be tempting to suggest that when the number of QE's is even ($N=4$, 
8, 12) for $2N < 2l+1$, that the pseudopotential of the Fermion pairs (FP's) 
would be subharmonic at $\nu_{\rm FP} =1/5$ (corresponding to $\nu_{\rm QE}=
1/2$), and that the Fermion pairs would themselves form pairs.
Then, only values of $N$ divisible by four would lead to condensed states.
There are two problems with this hypothesis.
The first is that the relation between $2l$ and $N$ would change from the 
values $2l=2N-3$ and $2N+1$ found numerically.
The second is that we do not know the Fermion pair--Fermion pair interaction
with a great degree of confidence.
These difficulties are being investigated, but at the moment, they call into 
question the validity of our simple ``complete pairing'' model.
Despite this, we are confident from our numerical and analytical work that 
some pairing of the QP excitations must occur so that the QP's can avoid 
pair states with the maximum repulsion.

\noindent
The authors acknowledge partial support from the Material Sciences 
Program -- Basic Energy Sciences of the US Department of Energy through 
Grant DE-FG 02-97ER45657. 
AW acknowledges support from Grant 2P03B02424 of the Polish KBN and 
thanks V.J.~Goldman and R.G.~Mani for useful discussions. 
KSY acknowledges partial support from ABRL (R14-2002-029-01002-0) of 
the KOSEF.
We also thank Dr. Jennifer J. Quinn for helpful discussions.

\end{document}